\documentclass[intlimits,twoside,a4paper]{article}

\usepackage{amsmath,amssymb}

\usepackage{braket}
\usepackage{color}
\usepackage{ulem}

\usepackage[eqsecnum]{cmpj3}
\usepackage{bm}

\issue{2026}{29}{2}{23705}
\doinumber{10.5488/CMP.29.23705}
\title[Field-tunable quadruple-$Q$ states by frustration]%
{Field-tunable quadruple-$Q$ states driven by momentum-space frustration}
\author[S. Hayami]{S. Hayami\orcid{0000-0001-9186-6958}\thanks{Email: \email{hayami@phys.sci.hokudai.ac.jp}.}
}
\address{
Graduate School of Science, Hokkaido University, Sapporo 060-0810, Japan
}

\date{Received 13 April 2026; revised 08 May 2026; accepted 11 May 2026; published 29 June 2026}

\begin{document}

\maketitle

\begin{abstract}
Multiple-$Q$ magnetism in itinerant electron systems enables complex spin crystals and noncoplanar textures even in centrosymmetric settings. We study a minimal momentum-space spin model on a square lattice with four symmetry-related ordering wave vectors, including bilinear and biquadratic interactions under an out-of-plane magnetic field. Using simulated annealing, we obtain the field-dependent phase diagram and identify successive transitions among single-$Q$, double-$Q$, and multiple inequivalent quadruple-$Q$ states. The quadruple-$Q$ manifold exhibits rich internal structures: the states sharing the same wave vectors differ in phase locking, amplitude distribution, and noncoplanarity, leading to distinct real-space textures and scalar spin chirality patterns. Our results demonstrate that momentum-space frustration and biquadratic coupling provide an efficient route to stabilizing diverse quadruple-$Q$ spin crystals, offering a general framework for higher-order spin textures in centrosymmetric itinerant magnets.
\printkeywords
%
\end{abstract}

\section{Introduction}

Magnetic ordering in itinerant electron systems often exhibits a complexity that goes far beyond the
standard paradigm of a single ordering wave vector~\cite{Gruner_RevModPhys.66.1, Baltz_RevModPhys.90.015005}.
In metallic magnets, the effective spin interaction is inherently nonlocal and momentum dependent,
reflecting the structure of the underlying Fermi surface~\cite{Yambe_PhysRevB.106.174437}.
As a consequence, magnetic instabilities are frequently characterized not by one isolated peak in
momentum space, but rather by the coexistence of several nearly degenerate ordering wave-vector channels.
When multiple wave vectors compete on comparable footing, the system is no longer forced to select a unique spiral or density-wave modulation.
Instead, the magnetic order may emerge through coherent superpositions of several symmetry-related modes, giving rise to multiple-$Q$ states~\cite{Bak_PhysRevLett.40.800, McEwen_PhysRevB.34.1781, zochowski1986thermal, forgan1990magnetic, Longfield_PhysRevB.66.054417, Bernhoeft_PhysRevB.69.174415, stewart2004phase, Watson_PhysRevB.53.726, Harris_PhysRevB.74.134411, schweizer2008accurate, Szabo_PhysRevLett.129.247201}.
Such phases provide a versatile route to unconventional spin textures, including noncoplanar spin
crystals and topological configurations like skyrmions~\cite{skyrme1962unified, Bogdanov89, Bogdanov94, rossler2006spontaneous, Muhlbauer_2009skyrmion, Yi_PhysRevB.80.054416, yu2010real, Butenko_PhysRevB.82.052403, Munzer_PhysRevB.81.041203, adams2010skyrmion, yu2011near, heinze2011spontaneous, Adams_PhysRevLett.107.217206, nagaosa2013topological, Hayami_PhysRevB.105.014408}, even in centrosymmetric magnets where the conventional
Dzyaloshinskii--Moriya mechanism~\cite{dzyaloshinsky1958thermodynamic,moriya1960anisotropic} is absent. 
These noncoplanar magnets are of particular interest because they generate an emergent real-space Berry curvature, leading to unconventional transport phenomena such as the topological Hall effect~\cite{Nagaosa_RevModPhys.82.1539, Xiao_RevModPhys.82.1959,Ohgushi_PhysRevB.62.R6065,taguchi2001spin, tatara2002chirality, machida2007unconventional, Shindou_PhysRevLett.87.116801, Martin_PhysRevLett.101.156402, Neubauer_PhysRevLett.102.186602, takatsu2010unconventional,ueland2012controllable, Hamamoto_PhysRevB.92.115417,nakazawa2018topological} and nonlinear nonreciprocal transport~\cite{tokura2018nonreciprocal, Hayami_PhysRevResearch.3.043158, Hayami_PhysRevB.106.014420, Eto_PhysRevLett.129.017201}.

A key aspect of this physics is that itinerant magnets naturally generate higher-order effective spin
interactions beyond the bilinear exchange level. 
In particular, quartic contributions, which can be viewed as arising from the coupling between spin
degrees of freedom and itinerant electrons, play an essential role in selecting among competing
multi-component ordering tendencies~\cite{Martin_PhysRevLett.101.156402, Akagi_JPSJ.79.083711, Barros_PhysRevB.88.235101, Venderbos_PhysRevLett.108.126405, Jiang_PhysRevLett.114.216402, Venderbos_PhysRevB.93.115108, Barros_PhysRevB.90.245119, Ghosh_PhysRevB.93.024401, takagi2018multiple, Wang_PhysRevLett.124.207201, Saito_PhysRevB.108.094440}.
These higher-order terms may stabilize magnetic structures that cannot be obtained within purely
bilinear frustrated models, thereby enriching the landscape of possible modulated phases.
From an experimental perspective, complex multiple-$Q$ order has been reported in a growing class of centrosymmetric metallic compounds, especially in rare-earth intermetallics where long-range Ruderman-Kittel-Kasuya-Yosida (RKKY)-type exchange interactions~\cite{Ruderman, Kasuya, Yosida1957} and electronic itinerancy cooperate, such as Gd$_2$PdSi$_3$~\cite{Saha_PhysRevB.60.12162,kurumaji2019skyrmion, Hirschberger_PhysRevLett.125.076602, Nomoto_PhysRevLett.125.117204, Kumar_PhysRevB.101.144440,Spachmann_PhysRevB.103.184424}, GdRu$_2$Si$_2$~\cite{khanh2020nanometric, Matsuyama_PhysRevB.107.104421, Wood_PhysRevB.107.L180402, eremeev2023insight, Spethmann_PhysRevMaterials.8.064404, dong2025pseudogap}, Gd$_3$Ru$_4$Al$_{12}$~\cite{Nakamura_PhysRevB.98.054410, hirschberger2019skyrmion, Hirschberger_10.1088/1367-2630/abdef9, Nakamura_PhysRevB.107.014422, ogunbunmi2023magnetic}, Y$_3$Co$_8$Sn$_4$~\cite{takagi2018multiple}, MnSc$_2$S$_4$~\cite{Gao2016Spiral,gao2020fractional, Rosales_PhysRevB.105.224402, takeda2024magnon}, and SrFeO$_3$~\cite{Ishiwata_PhysRevB.84.054427, Ishiwata_PhysRevB.101.134406,Rogge_PhysRevMaterials.3.084404, Onose_PhysRevMaterials.4.114420}.
Field-induced skyrmion crystals and related modulated phases observed in such materials demonstrate that nontrivial spin crystallization can arise even without strong relativistic chiral interactions. 
These findings suggest that centrosymmetric itinerant magnets provide a fertile ground for exploring
multiple wave-vector interference phenomena.

In this context, the square lattice offers a particularly transparent platform.
Since several symmetry-related ordering wave vectors may become energetically competitive, the square-lattice geometry naturally supports a variety of multiple-$Q$ superpositions~\cite{Venderbos_PhysRevLett.109.166405, Solenov_PhysRevLett.108.096403, Shahzad_PhysRevB.96.224402, shahzad2017phase, Marcus_PhysRevLett.120.097201, Seo_PhysRevX.10.011035, singh2023transition}.
While double-$Q$ and triple-$Q$ states have been extensively discussed in connection with vortex
crystals and skyrmion phases, the systematic stabilization of quadruple-$Q$ order remains far less
understood. 
The quadruple-$Q$ case is especially intriguing because it represents a qualitatively distinct regime
of mode interference.
The involvement of four ordering wave-vector components allows for a much broader range of internal phase relationships, amplitude distributions, and real-space texture morphologies than lower-$Q$ superpositions.
This raises the possibility of realizing unconventional spin crystals.

Motivated by these considerations, the present work investigates multiple-$Q$ magnetism in a
centrosymmetric square-lattice model incorporating both momentum-space exchange competition and a biquadratic interaction term.
By means of simulated annealing calculations and detailed momentum-resolved diagnostics, we demonstrate that the interplay between momentum-space frustration and quartic spin couplings stabilizes a broad family of distinct quadruple-$Q$ states.
These phases include several noncoplanar spin crystals with characteristic real-space modulation
patterns and structure-factor fingerprints.
Our results establish quadruple-$Q$ magnetism as a natural extension of itinerant multiple-$Q$
physics and provide a systematic framework for identifying higher-order spin crystals in
centrosymmetric magnets.
They further suggest that square-lattice metallic systems, including rare-earth-based materials,
engineered thin films, and artificial superlattices with competing RKKY interactions, offer promising
routes toward realizing such unconventional quadruple-$Q$ ordered phases.

The rest of this paper is organized as follows. 
In section~\ref{sec: Model and method}, we introduce the momentum-space spin model on the square lattice and describe the effective bilinear and biquadratic interactions used in the present study. 
We also outline the numerical method based on simulated annealing employed to determine the magnetic ground states. 
In section~\ref{sec: Results}, we present the magnetic phase diagrams under an out-of-plane magnetic field and discuss the sequence of field-induced transitions among single-$Q$, double-$Q$, and quadruple-$Q$ phases. 
The result provides a detailed characterization of the quadruple-$Q$ states in terms of their real-space spin textures, scalar spin chirality distributions, and spin structure factors. 
Finally, section~\ref{sec: Conclusions} summarizes the main results and discusses their implications for stabilization of higher-order multiple-$Q$ spin crystals in centrosymmetric itinerant magnets.

\section{Model and method}
\label{sec: Model and method}

In this section, we introduce the effective spin model on a two-dimensional square lattice and describe the numerical procedure employed to determine the low-temperature magnetic phases. 
Our focus is on the stabilization of multiple-$Q$ and, in particular, quadruple-$Q$ states in centrosymmetric itinerant magnets driven by momentum-space frustration and higher-order interactions.

\subsection{Effective momentum-space spin model}

We consider an itinerant electron magnet on a two-dimensional square lattice with the spatial inversion symmetry, where conduction electrons interact with localized magnetic moments. 
In the weak-coupling regime, integrating out the itinerant electron degrees of freedom generates an effective spin model containing not only bilinear exchange interactions of the RKKY type but also higher-order multi-spin couplings originating from electronic processes beyond second order~\cite{Akagi_PhysRevLett.108.096401}.
Keeping the dominant low-energy contributions relevant to magnetic ordering instabilities, we adopt a minimal momentum-space description in which bilinear and biquadratic interactions compete under an external magnetic field~\cite{hayami2024stabilization}. 
The resulting effective Hamiltonian is written as
\begin{align}
\label{eq: Ham}
\mathcal{H} =
-2J\sum_{\nu=1}^{4}\bm{S}_{\bm{Q}_\nu}\cdot\bm{S}_{-\bm{Q}_\nu}
+\frac{2K}{N}\sum_{\nu=1}^{4}
\left(\bm{S}_{\bm{Q}_\nu}\cdot\bm{S}_{-\bm{Q}_\nu}\right)^2
-H\sum_i S_i^z ,
\end{align}
where $J$ denotes the effective bilinear interaction corresponding to the RKKY exchange, $K$ represents the biquadratic interaction arising from higher-order electronic processes, and $H$ is the magnetic field applied along the $z$ direction. 
The spin at site $i$ is treated as a classical unit vector, $|\bm{S}_i|=1$, and $N=L^2$ is the total number of lattice sites. 
For simplicity, we do not consider symmetry-allowed anisotropic interactions, such as Ising-type interaction including the single-ion anisotropy~\cite{leonov2015multiply, Hayami_PhysRevB.99.094420, hayami2020multiple, Hayami_PhysRevB.103.224418} and bond-dependent interaction~\cite{Michael_PhysRevB.91.155135,Lee_PhysRevB.91.064407,Lukas_PhysRevLett.117.277202,Rousochatzakis2016,yao2016topological, amoroso2020spontaneous, Hayami_doi:10.7566/JPSJ.89.103702, Hayami_PhysRevB.105.104428}, which can lead to the instability toward multiple-$Q$ states. 
Such effective wave-vector-based descriptions have shown to be useful in various magnetic materials, where dominant susceptibility peaks at characteristic ordering wave vectors govern the emergence of multiple-$Q$ states and topological spin textures, such as EuNiGe$_3$~\cite{Goetsch_PhysRevB.87.064406, Fabreges_PhysRevB.93.214414, singh2023transition,matsumura2024helicity} and EuPtSi~\cite{kakihana2018giant,kaneko2019unique,tabata2019magnetic, kakihana2019unique, Mishra_PhysRevB.100.125113, takeuchi2019magnetic, hayami2021field, Matsumura_PhysRevB.109.174437}.

\subsection{Ordering wave vectors and multiple-$Q$ competition}

\begin{figure}[h]
	\begin{center}
		\includegraphics[width=8.0 cm]{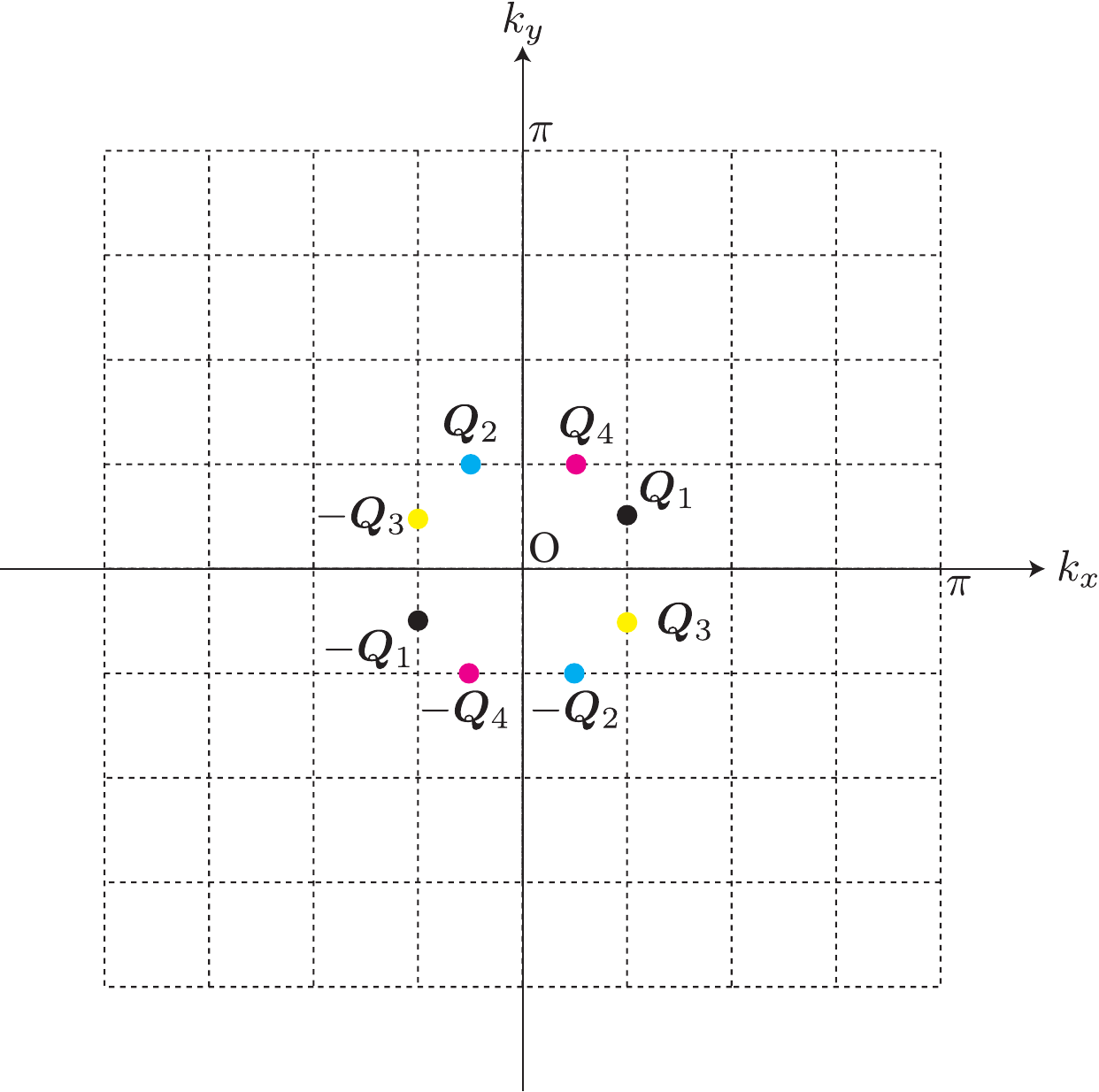}
		\caption{(Colour online) The four ordering wave vectors $\bm{Q}_1$--$\bm{Q}_4$ are illustrated in the Brillouin zone.
			They are generated by applying successive fourfold rotational and/or mirror operations to a representative wave vector.}\label{fig: Qvec}
	\end{center}
\end{figure}

A central feature of the present square-lattice model is the coexistence of several symmetry-related ordering wave-vector channels in momentum space. 
Instead of assuming a single dominant instability, we focus on four symmetry-related ordering wave vectors,
\begin{align}
&\pm\bm{Q}_1 =\pm(\piup/4, \piup/8),\quad
\pm\bm{Q}_2 =\pm(-\piup/8, \piup/4),\quad \nonumber \\
&\pm\bm{Q}_3 =\pm(\piup/4, -\piup/8),\quad
\pm\bm{Q}_4 =\pm(\piup/8, \piup/4),
\end{align}
which are connected by the fourfold rotational and mirror symmetries of the square-lattice Brillouin zone, as illustrated in figure~\ref{fig: Qvec}.

The presence of these four competing modes provides a natural basis for stabilizing not only single-$Q$ and double-$Q$ states but also a variety of unconventional quadruple-$Q$ superpositions. 
Such a reciprocal-space competition may be regarded as a form of momentum-space frustration, where several ordering tendencies remain energetically comparable and the system lowers its energy through coherent multiple-$Q$ interference~\cite{hayami2022multiple, Hayami_PhysRevB.105.174437}.
This framework is particularly relevant to centrosymmetric itinerant magnets, where complex spin crystal phases can emerge even without relativistic DM interactions, driven instead by higher-order couplings inherent to itinerant electron systems.

\subsection{Numerical procedure}

To determine the low-temperature magnetic phases of equation~(\ref{eq: Ham}), we perform simulated annealing simulations based on the standard single-spin-flip Metropolis algorithm. 
We consider a system of size $N=L^2$ with $L=16$ under periodic boundary conditions.
We adopt relatively small commensurate system sizes in order to stabilize the competing multiple-$Q$ states and avoid metastable configurations associated with nearly degenerate wave-vector combinations.

Starting from a high-temperature disordered configuration, the temperature is gradually reduced down to the low-temperature regime ($T/J=10^{-2}$), ensuring equilibration at each step by performing $10^{5}$--$10^{6}$ Monte Carlo sweeps. 
This procedure enables us to systematically map out the phase diagram and identify distinct multiple-$Q$ states, including several inequivalent quadruple-$Q$ spin textures.
We have also confirmed that the qualitative features of the phase diagram and the characteristic real-space textures remain unchanged for larger system sizes.

\subsection{Observables}

The magnetic structures obtained from the simulations are characterized through the spin structure factor,
\begin{align}
S_s^{\alpha}(\bm{q})=
\frac{1}{N}\sum_{i,j}
S_i^{\alpha}S_j^{\alpha}
\re^{i\bm{q}\cdot(\bm{r}_i-\bm{r}_j)},
\end{align}
where $\alpha=x,y,z$. 
We define the in-plane contribution as $S_s^{xy}(\bm{q})=S_s^{x}(\bm{q})+S_s^{y}(\bm{q})$, and the total component as $S_s(\bm{q})=S_s^{x}(\bm{q})+S_s^{y}(\bm{q})+S_s^{z}(\bm{q})$.
The momentum-resolved magnetic moment is given by
\begin{align}
m_{\bm{q}}=\sqrt{\frac{S_s(\bm{q})}{N}},
\end{align}
which provides a direct measure of how spectral weight is distributed among the competing ordering wave vectors.
We also compute the uniform magnetization along the field direction,
\begin{align}
M^z=\frac{1}{N}\sum_i S_i^z .
\end{align}

To quantify the degree of noncoplanarity and possible topological character of the spin textures, we evaluate the scalar spin chirality,
\begin{align}
\chi^{\mathrm{sc}} = \frac{1}{N}\sum_i \sum_{\delta=\pm1} 
\bm{S}_i \cdot \bigl(\bm{S}_{i+\delta\hat{x}} \times \bm{S}_{i+\delta\hat{y}}\bigr),
\end{align}
where $\hat{x}$ and $\hat{y}$ are unit basis vectors pointing along the Cartesian $x$ and $y$ axes, respectively.
A finite scalar spin chirality signals the emergence of nontrivial multiple-$Q$ interference effects and provides an indicator of topologically noncoplanar spin configurations.

\section{Results}
\label{sec: Results}

In this section, we present the magnetic phase diagram of the square-lattice momentum-space model in equation~(\ref{eq: Ham}) and discuss the evolution of multiple-$Q$ states as a function of the biquadratic interaction $K$ and the magnetic field $H$. 
By performing simulated annealing calculations, we identify a rich sequence of modulated phases, including several distinct quadruple-$Q$ spin crystals, as discussed in section~\ref{sec: Magnetic phase diagram}. 
Representative real-space spin textures, scalar spin chirality patterns, and momentum-resolved structure factors are summarized in sections~\ref{sec: Weak biquadratic coupling regime}--\ref{sec: Very strong biquadratic coupling regime}.

\subsection{Magnetic phase diagram}
\label{sec: Magnetic phase diagram}

\begin{figure}[h]
	\begin{center}
		\includegraphics[scale=0.6]{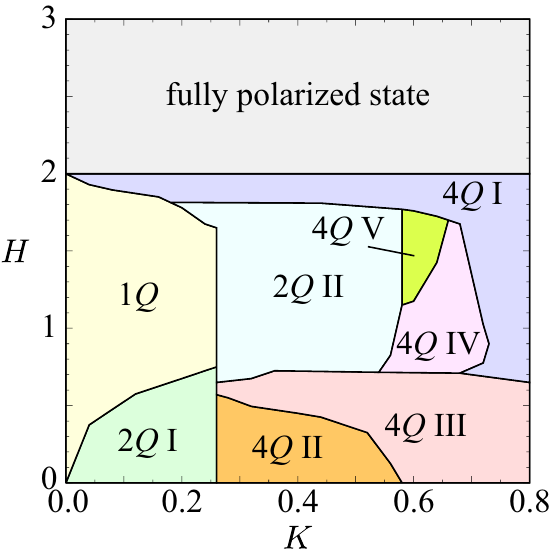}
		\caption{(Colour online) Magnetic phase diagram obtained for the spin model defined in equation~(\ref{eq: Ham}) in the low-temperature limit.
			The diagram is plotted in the parameter space spanned by the biquadratic interaction $K$, along the horizontal axis, and the applied magnetic field $H$ along the vertical axis.
			We fix the remaining parameter to $J=1$.
			The labels 1$Q$, 2$Q$, and 4$Q$ denote the single-$Q$ ordered phase, the double-$Q$ modulated phase, and the quadruple-$Q$ modulated phase, respectively. 
			The Roman numerals appended to the phase labels are used to distinguish between inequivalent phases of the same type.}
		\label{fig: PD}
	\end{center}
\end{figure}

We first provide an overview of the global magnetic phase diagram obtained in the low-temperature limit, which is summarized in figure~\ref{fig: PD}.
The phase diagram is mapped out in the $K$--$H$ plane, where the biquadratic interaction $K$ controls the strength of higher-order itinerant-electron-mediated couplings, while the magnetic field $H$ tunes the balance between modulated order and spin polarization.

As shown in figure~\ref{fig: PD}, the system exhibits a rich sequence of field-induced transitions among distinct modulated phases.
At weak $K$, the phase diagram is dominated by conventional single-$Q$ and double-$Q$ states,
which are continuously connected to spiral or vortex-like textures.
However, upon increasing $K$, the stability region of multiple-$Q$ phases expands rapidly, and a cascade of inequivalent quadruple-$Q$ states emerges over a wide range of magnetic fields.

A central feature of this diagram is that several different quadruple-$Q$ phases appear despite sharing the same set of ordering wave vectors $\bm{Q}_{1}$--$\bm{Q}_{4}$.
The Roman numerals are introduced to distinguish these distinct realizations within the same multiple-$Q$ class.
They reflect differences in the internal phase locking between the four wave-vector modulations, the relative distribution of spectral weight among symmetry-related wave-vector channels, and the resulting degree of noncoplanarity or topological character in real space. 
In this sense, the quadruple-$Q$ manifold realized here represents a qualitatively broader family of spin crystals than the well-studied double-$Q$ vortex states.
Thus, the phase diagram which demonstrates that momentum-space frustration among the four competing ordering wave-vector channels, combined with the biquadratic interaction, provides an efficient mechanism for stabilizing an extensive variety of unconventional quadruple-$Q$ spin textures in centrosymmetric square-lattice magnets.

In the following subsections, we examine the characteristic field evolution at several representative values of $K$,
and discuss how the magnetic textures, scalar spin chirality patterns, and spin structure factors  evolve across the phase boundaries.

\subsection{Weak biquadratic coupling regime}
\label{sec: Weak biquadratic coupling regime}

We begin with the weak-$K$ regime, where the bilinear RKKY-type exchange interaction provides the primary driving force for magnetic ordering.
In this regime, the system is expected to favor relatively simple modulated states selected mainly by the dominant susceptibility peaks at the ordering wave vectors,
while the biquadratic term plays only a secondary role in shaping the detailed interference pattern.
Nevertheless, even a small quartic contribution can become important once several ordering wave-vector channels remain nearly degenerate under an applied field.

\begin{figure}[h]
	\begin{center}
		\includegraphics[width=14.0 cm]{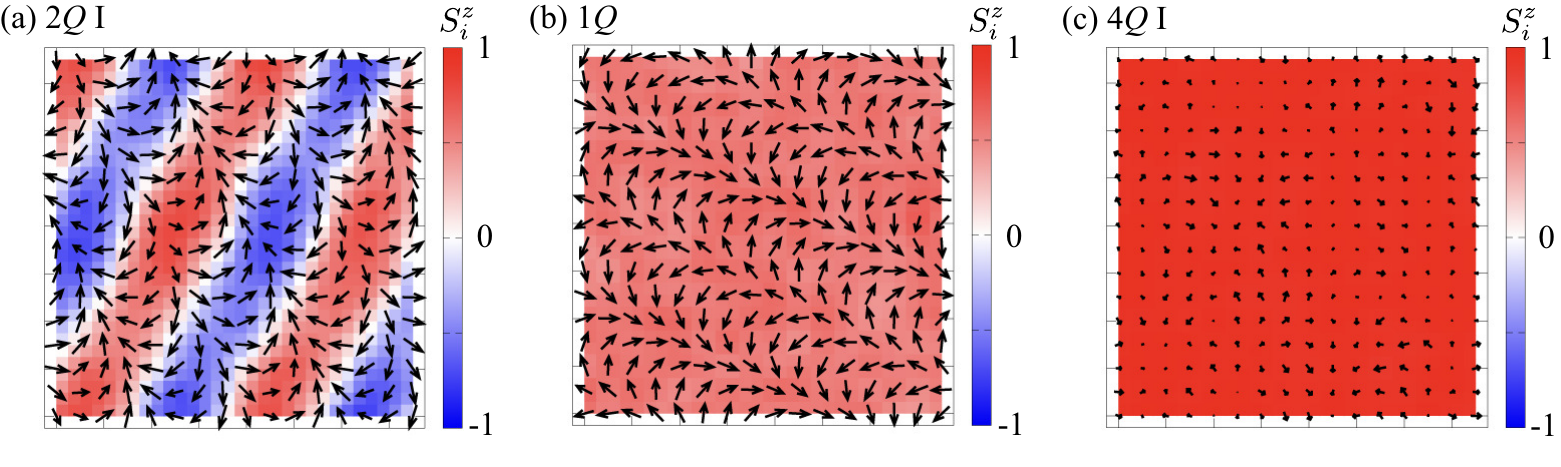}
		\caption{(Colour online) Representative real-space spin textures obtained by simulated annealing at $K=0.12$ are displayed.
			Panel (a) shows the double-$Q$ I (2$Q$ I) state at $H=0.1$, panel (b) corresponds to the single-$Q$ (1$Q$) state at $H=1$, and panel (c) represents the quadruple-$Q$ I (4$Q$ I) state at $H=1.9$.
			Spin orientations are indicated by arrows, and the color scale represents the out-of-plane component $S^z_i$.}
			\label{fig: Spin_K=0.12}
	\end{center}
\end{figure}

Figure~\ref{fig: Spin_K=0.12} summarizes representative real-space spin configurations stabilized at $K=0.12$.
At low magnetic field ($H=0.1$), the ground state is identified as the double-$Q$ I phase
[figure~\ref{fig: Spin_K=0.12}(a)], characterized by the coexistence of two symmetry-related modulation vectors.
The resulting spin texture exhibits an interference pattern distinct from a simple helical spiral, and already reflects the underlying momentum-space competition inherent to the square lattice.
Upon increasing $H$, the system undergoes a transition into a single-$Q$ phase
[figure~\ref{fig: Spin_K=0.12}(b)], in which one ordering wave-vector component becomes dominant.
This indicates that the Zeeman coupling favors the single-$Q$ conical structure with the in-plane oscillation.
Such a field-induced reduction of double-$Q$ instability is typical when the bilinear exchange dominates the energetics.

Interestingly, at higher magnetic field ($H=1.9$), the system enters a quadruple-$Q$ I state
[figure~\ref{fig: Spin_K=0.12}(c)].
Although the biquadratic interaction remains weak, the partial field polarization suppresses transverse fluctuations and enhances the relative importance of higher-order mode coupling.
As a result, coherent interference among all four ordering wave vectors becomes energetically favorable, demonstrating that even in the weak-$K$ regime, higher-$Q$ spin crystals can be activated by the magnetic field.

\begin{figure}[ht!]
\begin{center}
\includegraphics[width=5.0 cm]{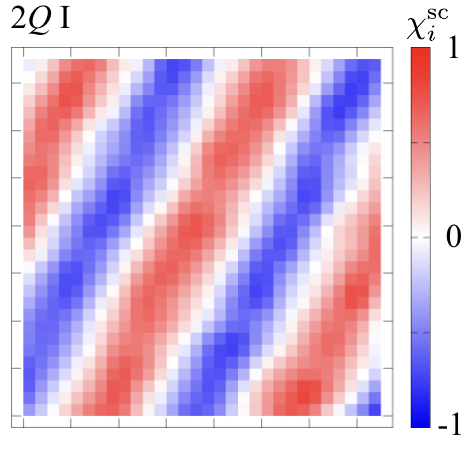}
\caption{(Colour online) Representative real-space patterns of the scalar spin chirality, obtained from simulated annealing calculations at $K=0.12$, are displayed. 
The data are for the double-$Q$ I (2$Q$ I) state at $H=0.1$.}
\label{fig: Chirality_K=0.12}
\end{center}
\end{figure}

The corresponding scalar spin chirality distribution in the double-$Q$ I phase is shown in figure~\ref{fig: Chirality_K=0.12}.
Whereas the spin texture is noncoplanar, the scalar spin chirality does not form a uniform skyrmion-like background.
Instead, it develops a striped modulation pattern, implying that the noncoplanarity is spatially oscillating and does not lead to a robust topological winding number~\cite{Hayami_PhysRevB.94.174420}.
This highlights that multiple-$Q$ superposition alone does not necessarily ensure a topologically nontrivial phase, and that the internal phase locking between modes plays a crucial role.

\begin{figure}[h]
	\begin{center}
		\includegraphics[width=10.0 cm]{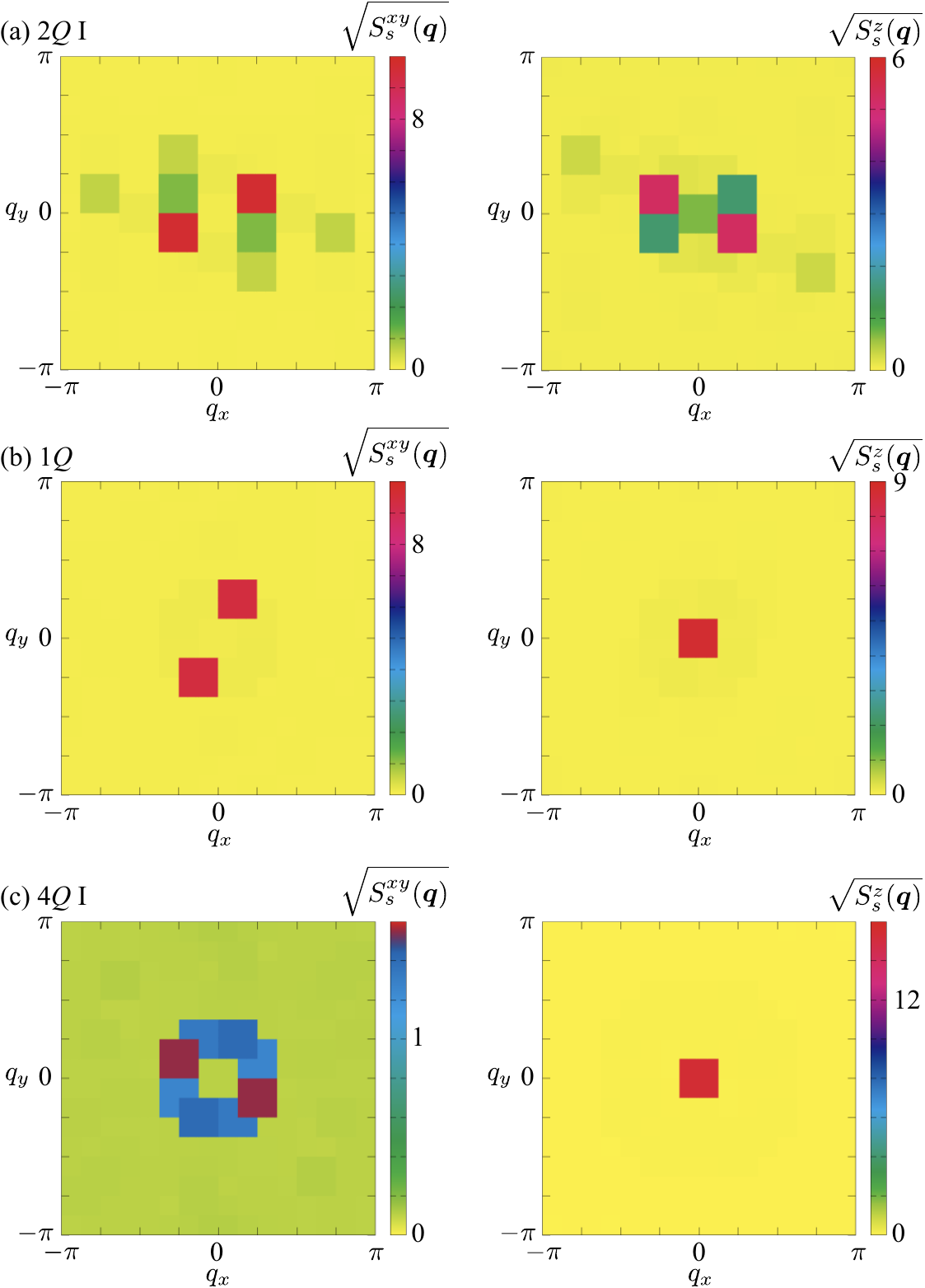}
		\caption{(Colour online) Momentum-space profiles of the spin structure factor are summarized by plotting $\scriptstyle \sqrt{S^{xy}_s(\bm q)}$ and $\scriptstyle \sqrt{S^{z}_s(\bm q)}$ for the phases in figure~\ref{fig: Spin_K=0.12}.
			The results are shown for (a) the double-$Q$ I (2$Q$ I) state at $H=0.1$, (b) the single-$Q$ (1$Q$) state at $H=1$, and (c) the quadruple-$Q$ I (4$Q$ I) state at $H=1.9$.
			In each case, the in-plane component appears in the left-hand panel, whereas the out-of-plane contribution is presented in the right-hand panel.  
			In the left-hand panel of (c), the color scale is adjusted to enhance the visibility of the inequivalent peaks.}
		\label{fig: Sq_K=0.12}
	\end{center}
\end{figure}

Momentum-space diagnostics further clarify these distinctions.
As shown in figure~\ref{fig: Sq_K=0.12}(a), the double-$Q$ I phase exhibits pronounced in-plane peaks concentrated at two ordering wave vectors, consistent with its two-mode character.
In the single-$Q$ phase, spectral weight collapses into a single dominant contribution
[figure~\ref{fig: Sq_K=0.12}(b)], reflecting the field-driven selection of one modulation channel.
By contrast, the quadruple-$Q$ I phase displays a much broader distribution of intensity over multiple $\bm{Q}_\nu$ components [figure~\ref{fig: Sq_K=0.12}(c)], providing clear evidence for a coherent four-mode superposition stabilized at high magnetic field.

\begin{figure}[h]
	\begin{center}
		\includegraphics[scale=0.5]{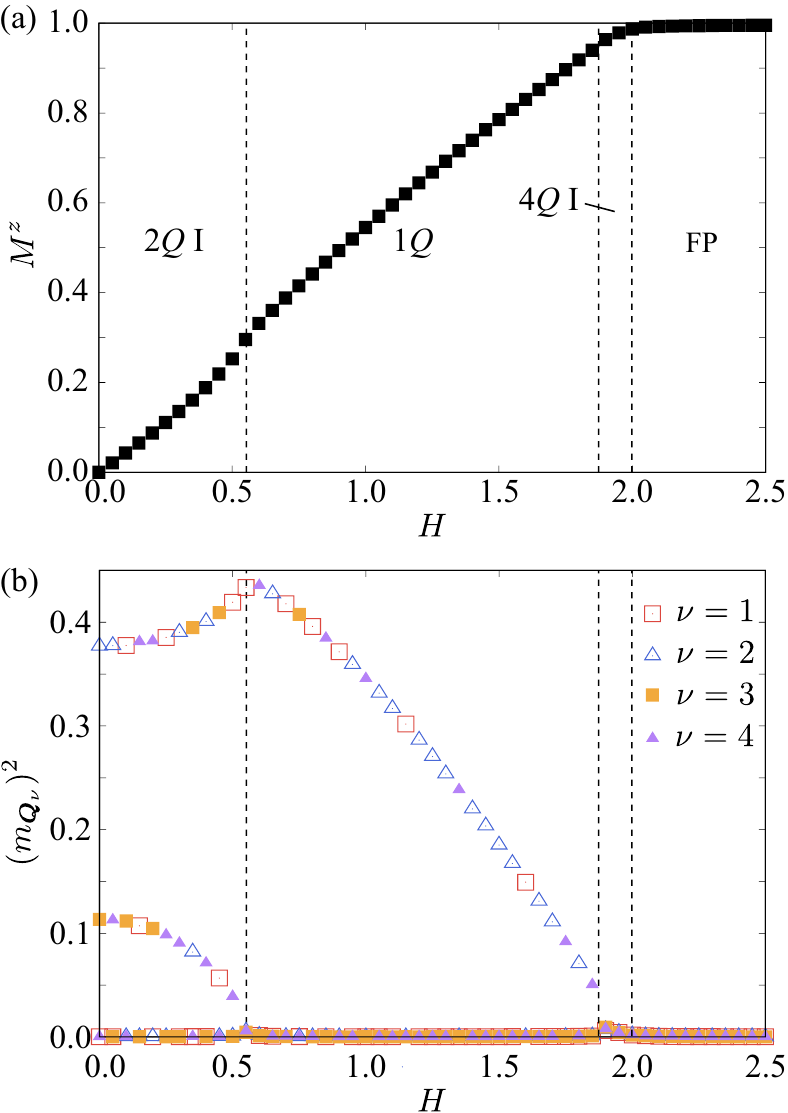}
		\caption{(Colour online) Magnetic-field evolution of (a) the magnetization and (b) the squared magnetic moments $(m_{\bm Q_\nu})^2$ at $K=0.12$.
			Vertical dashed lines indicate the phase transition points separating different magnetic states.}
		\label{mag_K=0.12}
	\end{center}
\end{figure}

Finally, the field evolution of the uniform magnetization and the squared Fourier moments is summarized in figure~\ref{mag_K=0.12}. 
The magnetization in figure~\ref{mag_K=0.12}(a) increases smoothly with field, reflecting the progressive canting of spins toward the field direction.
The momentum-resolved quantities $(m_{\bm{Q}_\nu})^2$ in figure~\ref{mag_K=0.12}(b) also evolve continuously across the transition from the double-$Q$ I phase to the intermediate single-$Q$ phase, indicating that this phase change is accompanied by a gradual redistribution of spectral weight among the ordering wave-vector channels.
The apparent switching between different components of $(m_{\bm{Q}_\nu})^2$ in the same phase originates from the random initial spin configurations employed in the simulated annealing procedure. 
Since the resulting magnetic states are connected by the symmetry operations of the square lattice, they are energetically degenerate and correspond to symmetry-equivalent magnetic domains rather than distinct magnetic phases.
By contrast, a clear jump in $(m_{\bm{Q}_\nu})^2$ is observed at the boundary between the single-$Q$ and quadruple-$Q$ I phases.
This discontinuity signals an abrupt reconstruction of the internal mode composition, where the system switches from a single-component modulation to a coherent quadruple-$Q$ superposition.
Therefore, even in the weak biquadratic coupling regime, the applied magnetic field can drive both continuous and discontinuous reorganizations of multiple-$Q$ order through momentum-space redistribution.

\subsection{Intermediate biquadratic coupling regime}

We next consider an intermediate coupling strength, $K=0.4$, where the biquadratic interaction becomes sufficiently strong to compete with the bilinear RKKY term already in the low-field region.
In this regime, quartic mode coupling no longer acts as a small correction but instead plays an active role in selecting coherent superpositions among the four symmetry-related ordering wave-vector channels.
As a result, quadruple-$Q$ order emerges as a central organizing principle of the phase diagram.

\begin{figure}[h]
	\begin{center}
		\includegraphics[width=14.0 cm]{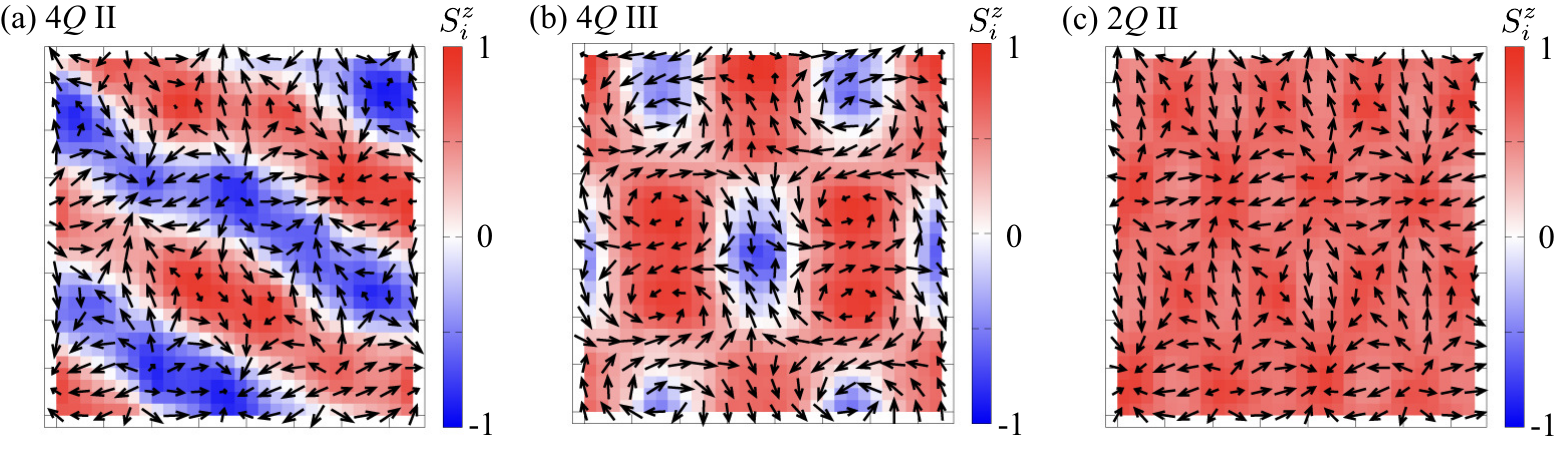}
		\caption{(Colour online) Representative real-space spin textures obtained by simulated annealing at $K=0.4$ are displayed.
			Panel (a) shows the quadruple-$Q$ II (4$Q$ II) state at $H=0.1$, panel (b) corresponds to the quadruple-$Q$ III (4$Q$ III) state at $H=0.6$, and panel (c) represents the double-$Q$ II (2$Q$ II) state at $H=1$.
			Spin orientations are indicated by arrows, and the color scale represents the out-of-plane component $S^z_i$.}
		\label{fig: Spin_K=0.40}
	\end{center}
\end{figure}

Representative real-space spin textures stabilized at $K=0.4$ are displayed in figure~\ref{fig: Spin_K=0.40}.
At very low magnetic field ($H=0.1$), the system realizes the quadruple-$Q$ II phase [figure~\ref{fig: Spin_K=0.40}(a)].
This state exhibits a strongly noncoplanar modulation involving all four ordering wave vectors, leading to a complex spin-crystal pattern that cannot be reduced to a simple vortex lattice. 
The enhanced stability of such a quadruple-$Q$ phase already at a weak field highlights the fact that
intermediate biquadratic coupling efficiently promotes higher-order interference among competing modes.

Upon increasing the magnetic field to $H=0.6$, the system undergoes a transition into another quadruple-$Q$ phase, labeled $4Q$ III [figure~\ref{fig: Spin_K=0.40}(b)].
Although both phases involve four ordering wave-vector components, their internal structures differ substantially.
In particular, the relative phase locking and amplitude distribution among the $\bm{Q}_\nu$ channels are reorganized, producing a distinct real-space modulation pattern.
This demonstrates that the square-lattice model supports multiple inequivalent realizations of quadruple-$Q$ order, even within the same $Q$-multiplicity class.

At a higher magnetic field ($H=1$), the system enters the double-$Q$ II phase
[figure~\ref{fig: Spin_K=0.40}(c)].
The appearance of this phase indicates that, even at intermediate $K$, the competition among different multiple-$Q$ combinations remains delicate.
The field partially suppresses some ordering wave-vector channels while favoring others, resulting in a reduced-$Q$ superposition in this regime.
Upon further increasing $H$, however, the system does not evolve directly into the trivial polarized phase.
Instead, an additional reconstruction takes place, and the quadruple-$Q$ I state becomes stabilized in a higher-field window.
This reentrance of a quadruple-$Q$ superposition highlights that the biquadratic interaction continues to promote multi-mode interference even under substantial Zeeman polarization.
Finally, at sufficiently strong magnetic fields, all modulations are suppressed and the spins become almost uniformly aligned, leading to the fully polarized state.

\begin{figure}[h]
	\begin{center}
		\includegraphics[width=14.0 cm]{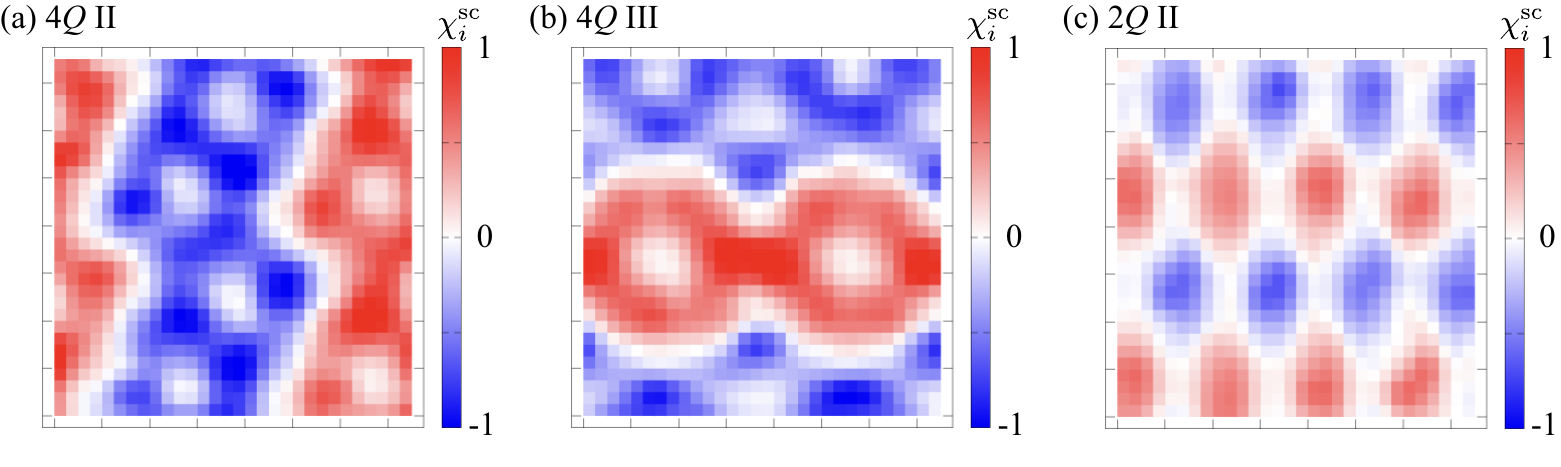}
		\caption{(Colour online) Representative real-space patterns of the scalar spin chirality, obtained from simulated annealing calculations at $K=0.4$, are displayed.
			Panel (a) shows the quadruple-$Q$ II (4$Q$ II) state at $H=0.1$, panel (b) corresponds to the quadruple-$Q$ III (4$Q$ III) state at $H=0.6$, and panel (c) represents the double-$Q$ II (2$Q$ II) state at $H=1$.}
		\label{fig: Chirality_K=0.40}
	\end{center}
\end{figure}

The evolution of noncoplanarity is further clarified by the scalar spin chirality distributions shown in figure~\ref{fig: Chirality_K=0.40}.
Both quadruple-$Q$ II and quadruple-$Q$ III phases generate pronounced scalar spin chirality modulations
[figures~\ref{fig: Chirality_K=0.40}(a) and \ref{fig: Chirality_K=0.40}(b)], reflecting their strongly three-dimensional spin arrangements.
The spatial pattern of scalar spin chirality differs between the two phases, providing an additional fingerprint of their inequivalence.
Interestingly, the double-$Q$ II phase also exhibits a highly nontrivial scalar spin chirality landscape rather than a simple suppression of scalar spin chirality.
As shown in figure~\ref{fig: Chirality_K=0.40}(c), the scalar spin chirality forms a complex spatial modulation with alternating positive and negative regions.
The resulting distribution bears a close resemblance to the characteristic pattern of a meron--antimeron crystal~\cite{brey1996skyrme, Ezawa_PhysRevB.83.100408, Lin_PhysRevB.91.224407, Tan_PhysRevB.94.014433, yu2018transformation, kurumaji2019skyrmion, Hayami_PhysRevB.104.094425}, suggesting that the double-$Q$ II state hosts an intricate noncoplanar texture with local topological motifs, even though the net scalar spin chirality is compensated within the magnetic unit cell.

\begin{figure}[h]
\begin{center}
\includegraphics[scale=0.5]{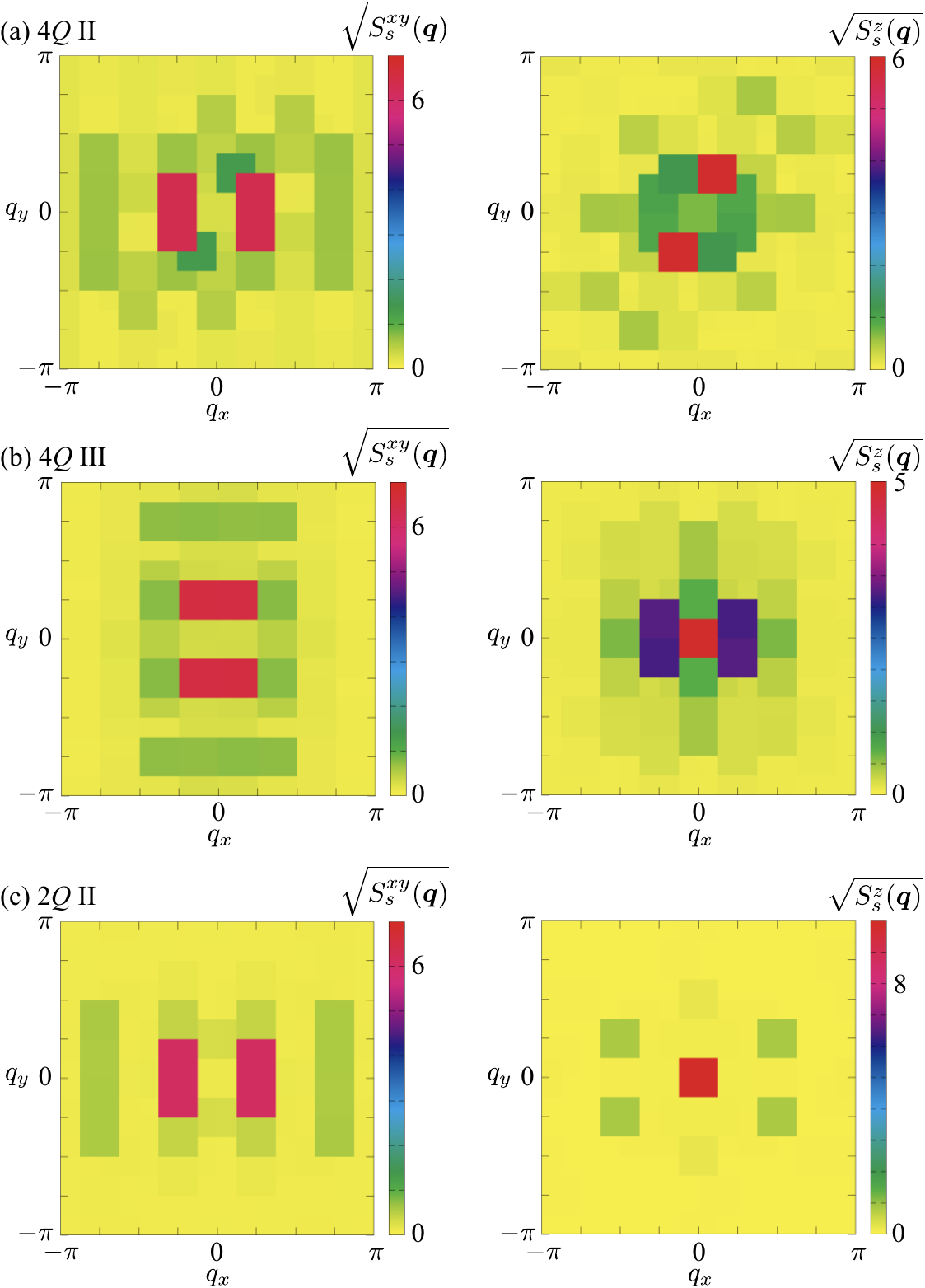}
\caption{(Colour online) Momentum-space profiles of the spin structure factor are summarized by plotting $\scriptstyle \sqrt{S^{xy}_s(\bm q)}$ and $\scriptstyle \sqrt{S^{z}_s(\bm q)}$ for the phases in figure~\ref{fig: Spin_K=0.40}.
The results are shown for (a) the quadruple-$Q$ II (4$Q$ II) state at $H=0.1$, (b) the quadruple-$Q$ III (4$Q$ III) state at $H=1$, and (c) the double-$Q$ II (2$Q$ II) state at $H=1.9$.
In each case, the in-plane component appears in the left-hand panel, whereas the out-of-plane contribution is presented in the right-hand panel.}
\label{fig: Sq_K=0.40}
\end{center}
\end{figure}   
\unskip

Momentum-space diagnostics are summarized in figure~\ref{fig: Sq_K=0.40}.
In the quadruple-$Q$ II phase, the spin structure factor shows substantial intensity distributed over multiple ordering wave vectors together with finite in-plane and out-of-plane contributions [figure~\ref{fig: Sq_K=0.40}(a)], confirming the coherent four-mode character of the state.
The quadruple-$Q$ III phase exhibits a qualitatively different redistribution of spectral weight [figure~\ref{fig: Sq_K=0.40}(b)], indicating that the transition between the two quadruple-$Q$ phases involves a genuine reconstruction of the internal mode composition, rather than a trivial symmetry-related deformation.
In the double-$Q$ II phase, the spectral weight becomes concentrated in fewer channels [figure~\ref{fig: Sq_K=0.40}(c)], reflecting the reduced multiplicity of the modulation.

\begin{figure}[h]
	\begin{center}
		\includegraphics[scale=0.5]{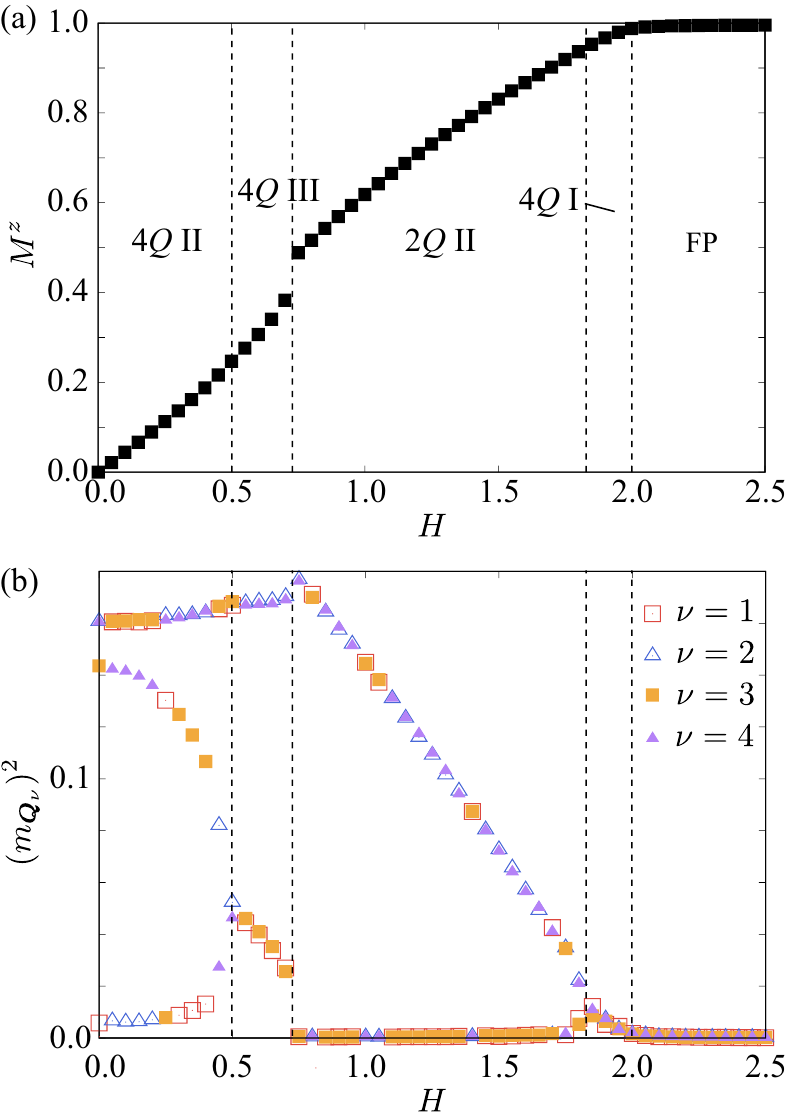}
		\caption{(Colour online) Magnetic-field evolution of (a) the magnetization and (b) the squared magnetic moments $(m_{\bm Q_\nu})^2$ at $K=0.4$.
			Vertical dashed lines indicate the phase transition points separating different magnetic states.}
		\label{mag_K=0.40}
	\end{center}
\end{figure}

Finally, the field dependence of the magnetization $M^z$ and the squared Fourier moments $(m_{\bm{Q}_\nu})^2$ is summarized in figure~\ref{mag_K=0.40}.
Whereas the magnetization in figure~\ref{mag_K=0.40}(a) increases smoothly with field, the momentum-resolved components provide a more sensitive probe of the underlying texture evolution.
In the low- and intermediate-field regions, $(m_{\bm{Q}_\nu})^2$ varies continuously across the crossover between the quadruple-$Q$ II and quadruple-$Q$ III phases, as shown in figure~\ref{mag_K=0.40}(b), indicating that the reorganization of the quadruple-$Q$
superposition proceeds without an abrupt redistribution of spectral weight.
By contrast, a clear jump in $(m_{\bm{Q}_\nu})^2$ is observed only at the transition between the quadruple-$Q$ III and double-$Q$ II phases, signaling a discontinuous change in the dominant ordering wave-vector channels and a more drastic reconstruction of the internal mode composition.
At higher magnetic fields, the system further evolves toward the nearly polarized regime through a gradual suppression of the finite-$Q$ components.

Thus, the intermediate-$K$ regime illustrates that biquadratic interactions can stabilize a hierarchy of
quadruple-$Q$ spin crystals, where some phase boundaries are characterized by smooth deformations of the multiple-$Q$ texture, while others involve abrupt mode-selective transitions.

\subsection{Strong biquadratic coupling regime}

We now turn to the strong biquadratic coupling regime, exemplified by $K=0.6$, where quartic spin interactions play a dominant role in shaping the magnetic ground states.
In this parameter range, the biquadratic term strongly enhances cooperative interference among all four ordering wave-vector channels, and the system is driven deep into a regime where quadruple-$Q$ superpositions become energetically favorable over lower-$Q$ alternatives.
As a consequence, several distinct realizations of quadruple-$Q$ spin crystals emerge in close proximity, highlighting the remarkable richness of the strong-$K$ landscape.

\begin{figure}[h]
	\begin{center}
		\includegraphics[width=10.0 cm]{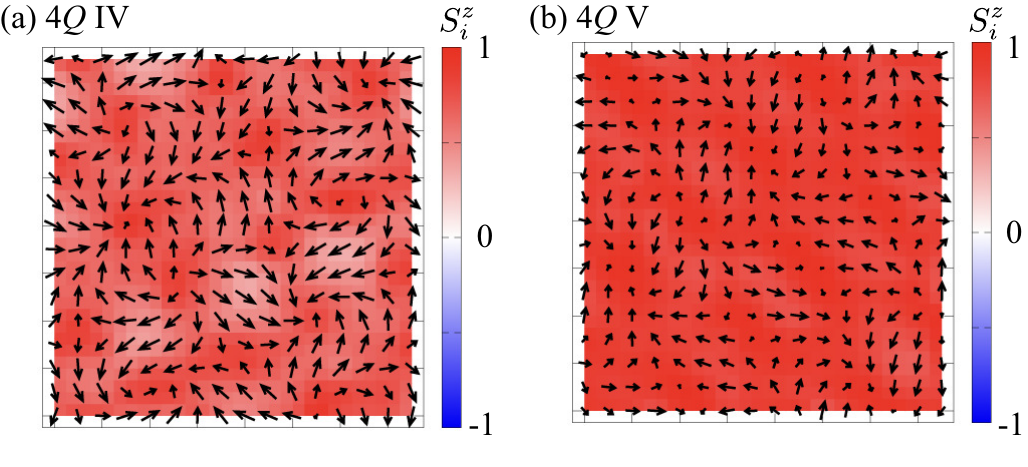}
		\caption{(Colour online) Representative real-space spin textures obtained by simulated annealing at $K=0.6$ are displayed.
			Panel (a) shows the quadruple-$Q$ IV (4$Q$ IV) state at $H=1$ and panel (b) corresponds to the quadruple-$Q$ V (4$Q$ V) state at $H=1.5$.
			Spin orientations are indicated by arrows, and the color scale represents the out-of-plane component $S^z_i$.}
		\label{fig: Spin_K=0.60}
	\end{center}
\end{figure}

Representative real-space spin configurations obtained at $K=0.6$ are presented in figure~\ref{fig: Spin_K=0.60}.
At $H=1$, the system stabilizes the quadruple-$Q$ IV phase [figure~\ref{fig: Spin_K=0.60}(a)].
This state exhibits a robust noncoplanar arrangement involving coherent contributions from all
four symmetry-related ordering wave vectors.
Compared with the double-$Q$ II phase found at intermediate $K$ for the same $H$, the present spin texture shows a more pronounced three-dimensional modulation of spin components, indicating that strong biquadratic coupling locks the multi-mode interference more rigidly and suppresses residual degeneracies among competing configurations.

Upon increasing the magnetic field to $H=1.5$, the system undergoes a transition into another
quadruple-$Q$ phase, labeled $4Q$ V [figure~\ref{fig: Spin_K=0.60}(b)].
Although this phase also belongs to the quadruple-$Q$ class, its internal structure is clearly distinct.
In particular, the field reorganizes the relative phase relationships and amplitude balance among the
$\bm{Q}_\nu$ components, producing a different spatial modulation pattern.
The close energetic competition between qusdruple-$Q$ IV and quadruple-$Q$ V phases demonstrates that, under strong quartic
coupling, the square-lattice model supports multiple inequivalent quadruple-$Q$ crystals separated by field-induced texture reconstructions rather than simple canting processes.

\begin{figure}[h]
\begin{center}
\includegraphics[width=10.0 cm]{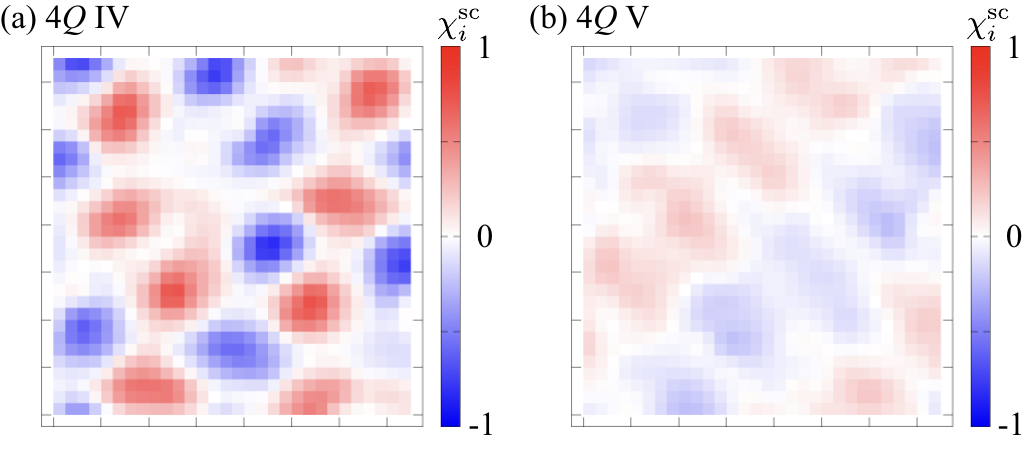}
\caption{(Colour online) Representative real-space patterns of the scalar spin chirality, obtained from simulated annealing calculations at $K=0.6$, are displayed.
Panel (a) shows the quadruple-$Q$ IV (4$Q$ IV) state at $H=1$ and panel (b) corresponds to the quadruple-$Q$ V (4$Q$ V) state at $H=1.5$.}
\label{fig: Chirality_K=0.60}
\end{center}
\end{figure}

The noncoplanar character of these phases is further reflected in the scalar spin chirality landscapes.
As shown in figures~\ref{fig: Chirality_K=0.60}(a) and \ref{fig: Chirality_K=0.60}(b), both quadruple-$Q$ IV and quadruple-$Q$ V states exhibit pronounced scalar spin chirality modulations, confirming the emergence of strongly three-dimensional spin arrangements.
The scalar spin chirality exhibits stripe-like modulation patterns with alternating positive and negative regions originating from the underlying quadruple-$Q$ interference. 
Although the finite-size cluster limits the spatial resolution of the textures, the modulation patterns remain spatially regular and consistent with the symmetry of the corresponding spin configurations. 
The scalar spin chirality, therefore, provides an important diagnostic of the quadruple-$Q$ states: while both phases carry sizable local scalar spin chirality, their distinct spatial organizations indicate that the transition between them involves a qualitative reconstruction of the underlying noncoplanar spin crystal rather than a smooth deformation.

\begin{figure}[h]
	\begin{center}
		\includegraphics[scale=0.55]{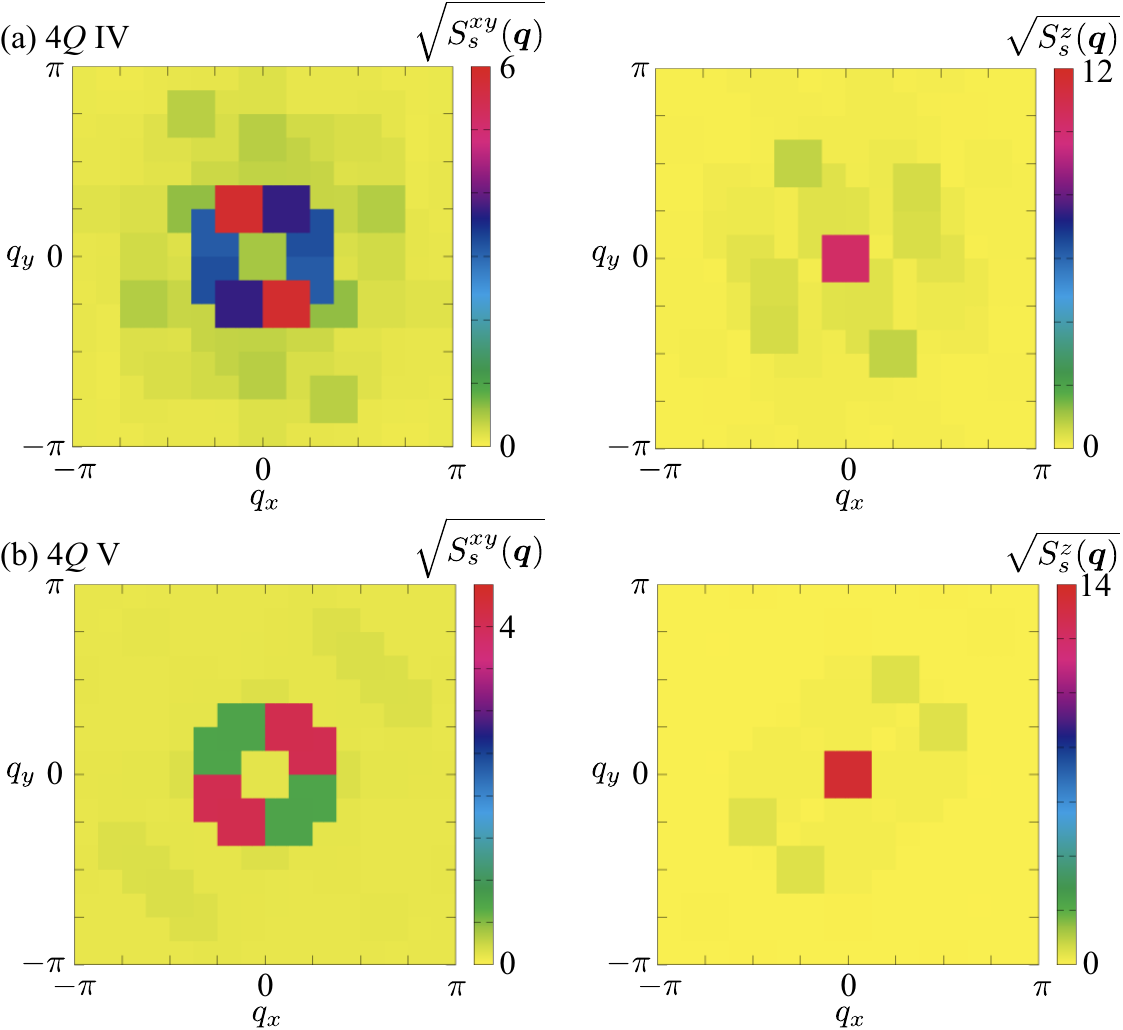}
		\caption{(Colour online) Momentum-space profiles of the spin structure factor are summarized by plotting $\scriptstyle \sqrt{S^{xy}_s(\bm q)}$ and $\scriptstyle \sqrt{S^{z}_s(\bm q)}$ for the phases in figure~\ref{fig: Spin_K=0.60}.
			The results are shown for (a) the quadruple-$Q$ IV (4$Q$ IV) state at $H=1$ and (b) the quadruple-$Q$ V (4$Q$ V) state at $H=1.5$.
			In each case, the in-plane component appears in the left-hand panel, whereas the out-of-plane contribution is presented in the right-hand panel.}
		\label{fig: Sq_K=0.60}
	\end{center}
\end{figure}

Momentum-space fingerprints of the strong-$K$ phases are summarized in figure~\ref{fig: Sq_K=0.60}.
Both the qu\-a\-dru\-p\-le-$Q$ IV and quadruple-$Q$ V states exhibit pronounced intensity at all four ordering
wave vectors, reflecting their common quadruple-$Q$ character
[figures~\ref{fig: Sq_K=0.60}(a) and \ref{fig: Sq_K=0.60}(b)].
The overall distribution of spectral weight is therefore qualitatively similar between the two phases,
indicating that they are built from comparable four-mode superpositions.
Nevertheless, closer inspection reveals subtle differences in the relative intensity balance among the
$\bm{Q}_\nu$ components.
Such quantitative changes suggest that the transition between quadruple-$Q$ IV and quadruple-$Q$ V states does not correspond to a simple symmetry operation, but rather to a reorganization of the internal phase locking and amplitude structure within the same quadruple-$Q$ manifold.

\begin{figure}[h]
	\begin{center}
		\includegraphics[scale=0.45]{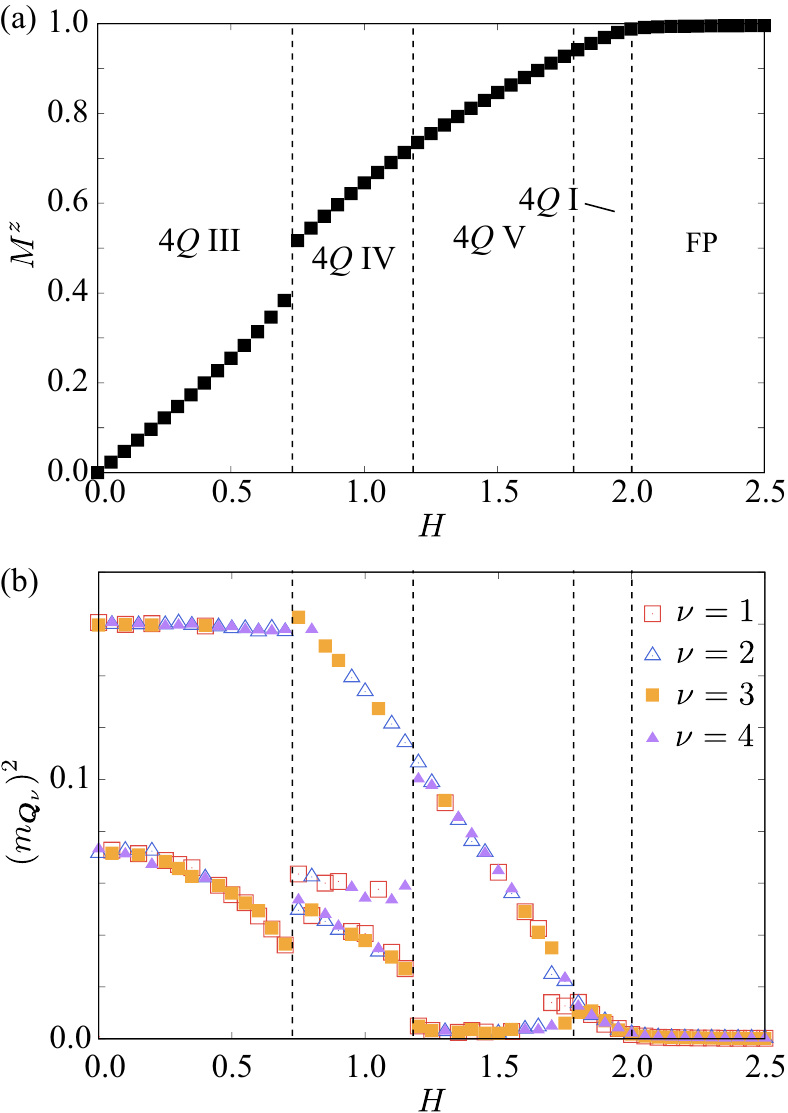}
		\caption{(Colour online) Magnetic-field evolution of (a) the magnetization and (b) the squared magnetic moments $(m_{\bm Q_\nu})^2$ at $K=0.6$.
			Vertical dashed lines indicate the phase transition points separating different magnetic states.}
		\label{mag_K=0.60}
	\end{center}
\end{figure}

Finally, the magnetization process and the evolution of squared Fourier moments at $K=0.6$
are summarized in figure~\ref{mag_K=0.60}.
While the net magnetization in figure~\ref{mag_K=0.60}(a) increases progressively with field, the momentum-resolved quantities $(m_{\bm{Q}_\nu})^2$ in figure~\ref{mag_K=0.60}(b) exhibit step-like redistributions across the transition fields.
This behavior indicates that the field evolution in the strong-$K$ regime is governed by successive reorganizations of the internal multiple-$Q$ composition, in which spectral weight is transferred nonuniformly among the competing ordering wave vectors before the system eventually approaches the polarized phase.
Overall, the results at $K=0.6$ establish that strong biquadratic coupling provides an efficient mechanism for stabilizing a hierarchy of robust and inequivalent quadruple-$Q$ spin crystals in centrosymmetric square-lattice itinerant magnets.

\subsection{Very strong biquadratic coupling regime}
\label{sec: Very strong biquadratic coupling regime}

Finally, we examine the very strong biquadratic coupling regime at $K=0.76$, which represents the case where quartic interactions overwhelmingly dominate the energetic balance.
In this regime, the biquadratic term strongly promotes cooperative interference among all four ordering wave-vector channels, and as a result, quadruple-$Q$ ordering becomes the most stable form of magnetic organization over a broad field range. 
Lower-$Q$ states are almost completely suppressed, indicating that the system is driven deep into a spin-crystal regime governed by higher-order mode locking rather than conventional bilinear exchange competition.

\begin{figure}[h]
	\begin{center}
		\includegraphics[scale=0.55]{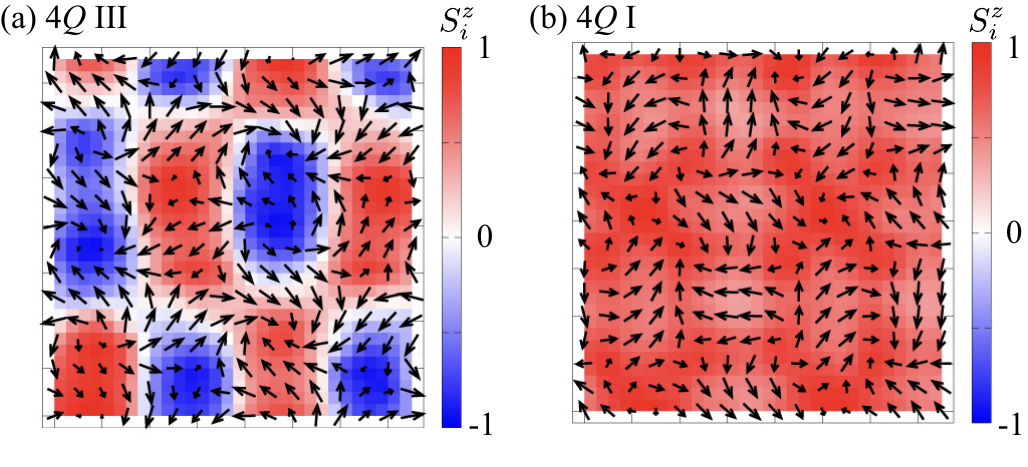}
		\caption{(Colour online) Representative real-space spin textures obtained by simulated annealing at $K=0.76$ are displayed.
			Panel (a) shows the quadruple-$Q$ III (4$Q$ III) state at $H=0.1$ and panel (b) corresponds to the quadruple-$Q$ I (4$Q$ I) state at $H=1$.
			Spin orientations are indicated by arrows, and the color scale represents the out-of-plane component $S^z_i$.}
		\label{fig: Spin_K=0.76}
	\end{center}
\end{figure}   

Representative real-space spin configurations obtained at $K=0.76$ are displayed in figure~\ref{fig: Spin_K=0.76}.
At very low field ($H=0.1$), the system stabilizes the quadruple-$Q$ III phase [figure~\ref{fig: Spin_K=0.76}(a)].
This state exhibits a strongly noncoplanar modulation pattern, in which the four ordering wave-vector
components interfere in a highly coherent manner.
The resulting texture forms a dense spin crystal with pronounced spatial variation of both in-plane
and out-of-plane spin components, reflecting the fact that very strong quartic coupling enforces
multi-mode coexistence already in the weak-field limit.

Upon increasing the field to $H=1$, the magnetic structure undergoes a qualitative reconstruction and
enters the quadruple-$Q$ I phase [figure~\ref{fig: Spin_K=0.76}(b)].
Although this phase belongs to the same quadruple-$Q$ multiplicity class, its internal arrangement is substantially different from that of quadruple-$Q$ III phase.
In particular, the applied field reshapes the balance between uniform polarization and modulated components, leading to a reorganization of relative phases and amplitudes among the $\bm{Q}_\nu$ channels.
This clearly demonstrates that, even when quadruple-$Q$ order remains dominant, the field can induce
distinct spin-crystal realizations through internal texture reconstruction rather than through a simple reduction of $Q$-multiplicity.

\begin{figure}[h]
	\begin{center}
		\includegraphics[scale=0.6]{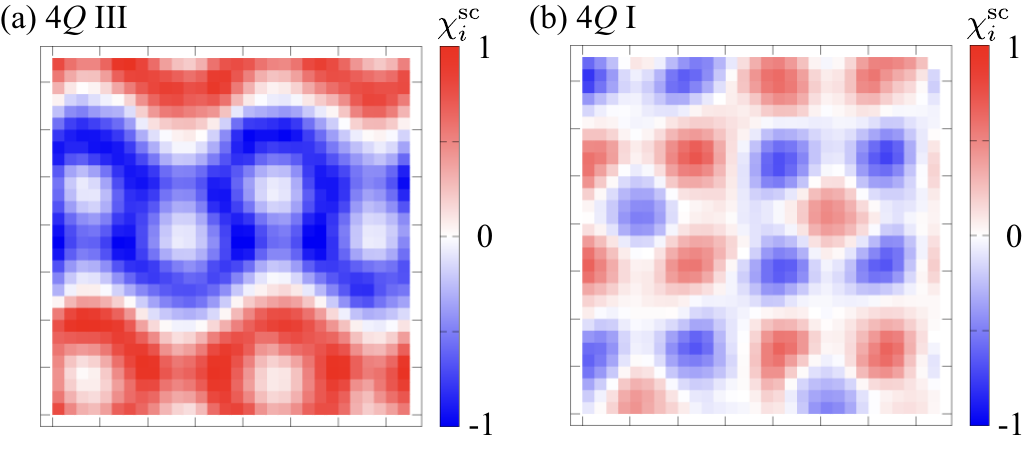}
		\caption{(Colour online) Representative real-space patterns of the scalar spin chirality, obtained from simulated annealing calculations at $K=0.76$, are displayed.
			Panel (a) shows the quadruple-$Q$ III (4$Q$ III) state at $H=0.1$ and panel (b) corresponds to the quadruple-$Q$ I (4$Q$ I) state at $H=1$.}
		\label{fig: Chirality_K=0.76}
	\end{center}
\end{figure}

The noncoplanar nature of these phases is further confirmed by the scalar spin chirality maps shown in
figure~\ref{fig: Chirality_K=0.76}.
Both the $4Q$ III and $4Q$ I states host sizable scalar spin chirality modulations, indicating robust
three-dimensional spin twisting, as shown in figures~\ref{fig: Chirality_K=0.76}(a) and \ref{fig: Chirality_K=0.76}(b).
Notably, the scalar spin chirality patterns differ in their spatial organization, suggesting that the transition
between the two phases involves a genuine rearrangement of the underlying topological spin geometry,
even though the overall quadruple-$Q$ character is preserved.

\begin{figure}[h]
	\begin{center}
		\includegraphics[scale=0.5]{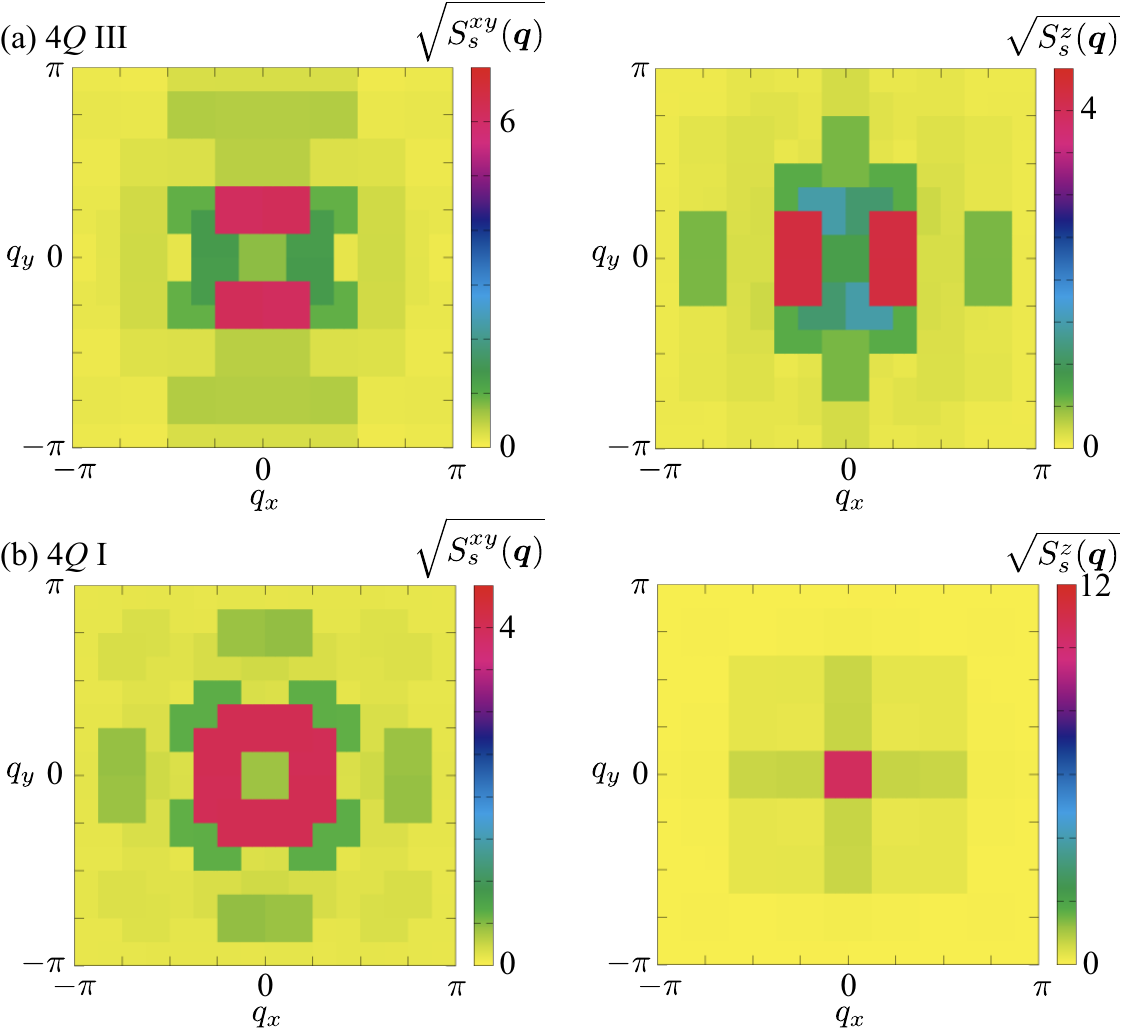}
		\caption{(Colour online) Momentum-space profiles of the spin structure factor are summarized by plotting $\scriptstyle \sqrt{S^{xy}_s(\bm q)}$ and $\scriptstyle \sqrt{S^{z}_s(\bm q)}$ for the phases in figure~\ref{fig: Spin_K=0.76}.
			The results are shown for (a) the quadruple-$Q$ III (4$Q$ III) state at $H=0.1$ and (b) the quadruple-$Q$ I (4$Q$ I) state at $H=1$.
			In each case, the in-plane component appears in the left-hand panel, whereas the out-of-plane contribution is presented in the right-hand panel.}
		\label{fig: Sq_K=0.76}
	\end{center}
\end{figure}

Momentum-space fingerprints of the very strong-$K$ regime are summarized in figure~\ref{fig: Sq_K=0.76}.
In the quadruple-$Q$ III phase, the structure factor exhibits a relatively broad distribution of
intensity among the four ordering wave vectors in both in-plane and out-of-plane components [figure~\ref{fig: Sq_K=0.76}(a)],
consistent with a strongly entangled multi-mode superposition.
By contrast, the quadruple-$Q$ I phase displays a more symmetric and sharply defined spectral pattern only in the in-plane component [figure~\ref{fig: Sq_K=0.76}(b)], reflecting a different internal locking of the ordering wave-vector components.
These reciprocal-space distinctions provide a clear evidence that multiple inequivalent quadruple-$Q$ spin crystals can be stabilized and tuned by a magnetic field even in the deep strong-coupling limit.

\begin{figure}[h!]
	\begin{center}
		\includegraphics[scale=0.45]{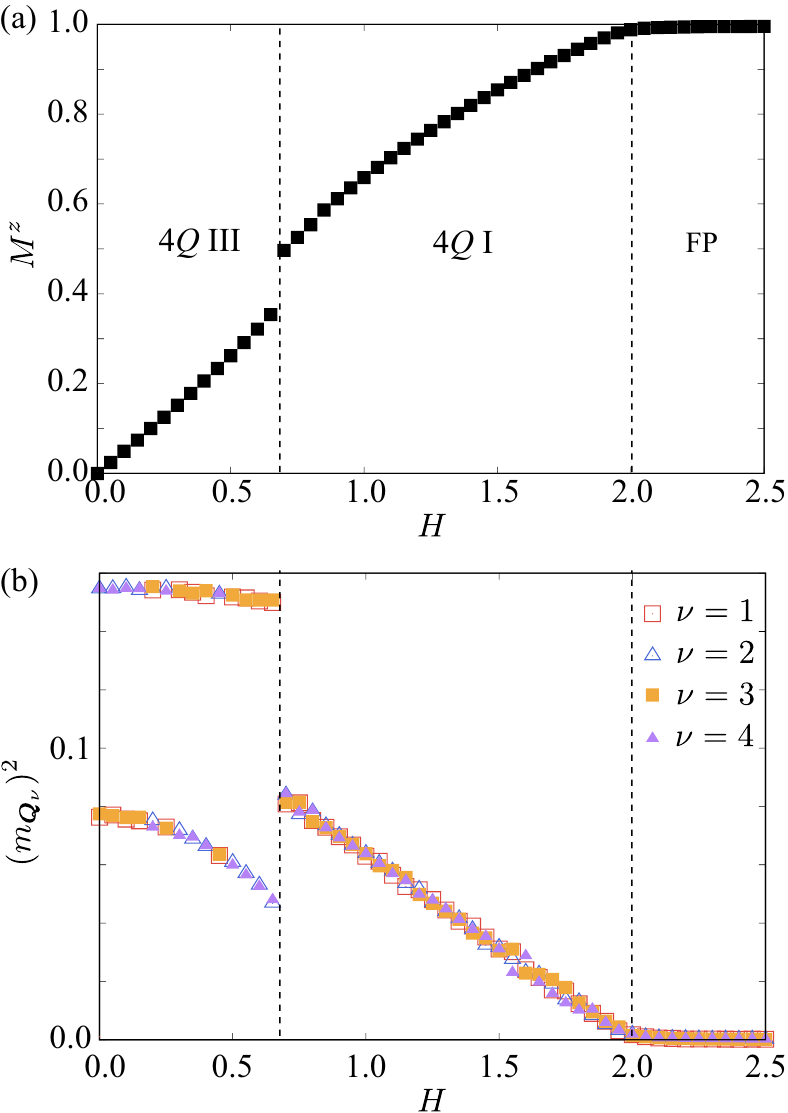}
		\caption{(Colour online) Magnetic-field evolution of (a) the magnetization and (b) the squared magnetic moments $(m_{\bm Q_\nu})^2$ at $K=0.76$.
			Vertical dashed lines indicate the phase transition points separating different magnetic states.}
		\label{mag_K=0.76}
	\end{center}
\end{figure}

Finally, the field evolution of the magnetization and squared Fourier moments at $K=0.76$ is shown in
figure~\ref{mag_K=0.76}.
The magnetization increases steadily with field, yet the quadruple-$Q$ order remains remarkably robust
up to relatively high fields before the eventual transition into the fully polarized phase, as shown in figure~\ref{mag_K=0.76}(a).
The momentum-resolved quantities $(m_{\bm{Q}_\nu})^2$ exhibit nonuniform and step-like changes across
the transition between the quadruple-$Q$ III and quadruple-$Q$ I states, as shown in figure~\ref{mag_K=0.76}(b), indicating that the field primarily drives internal redistributions among the four ordering wave-vector channels rather than immediately destroying the multiple-$Q$ interference.
Overall, these results establish that, in the very strong biquadratic coupling regime, centrosymmetric square-lattice itinerant magnets can host exceptionally stable and diverse quadruple-$Q$ spin crystals, whose internal structures remain highly tunable under external magnetic fields.

\section{Conclusions}
\label{sec: Conclusions}

We have studied field-induced multiple-$Q$ magnetism in a centrosymmetric square-lattice model that
captures a momentum-space competition among four symmetry-related ordering wave-vector channels.
By combining the effective bilinear interaction with a biquadratic term and performing simulated annealing in the low-temperature limit, we obtained a comprehensive magnetic phase diagram in the $K$--$H$ plane and clarified how the higher-order mode coupling reshapes the landscape of modulated phases.

Our main finding is the emergence of a broad and structured family of inequivalent quadruple-$Q$ states.
Although these phases are constructed from the same set of ordering wave vectors, they exhibit qualitatively different internal organizations, reflected in (i) distinct real-space spin crystals, (ii) characteristic scalar spin chirality patterns, and (iii) quantitative differences in the distribution of spectral weight among the $\bm{Q}_\nu$ channels.
The field evolution reveals that phase boundaries within the multiple-$Q$ manifold can be of different character.
In the weak biquadratic regime, the double-$Q$ I to single-$Q$ transition evolves smoothly, while the onset of the quadruple-$Q$ I phase is marked by an abrupt reconstruction. 
At intermediate biquadratic regime, the field induces several texture changes, with a clear discontinuity at the
boundary between quadruple-$Q$ III and double-$Q$ II phases.
For strong and very strong biquadratic regime, quadruple-$Q$ phases dominate a wide field range and remain robust
up to high fields before the fully polarized state is reached.

These results highlight a general physical message: in centrosymmetric itinerant magnets, momentum-space frustration among multiple nearly degenerate ordering wave-vector channels, when combined with higher-order itinerant-electron-mediated interactions, can stabilize higher-multiplicity spin crystals well beyond the conventional single-$Q$ and double-$Q$ scenarios.
The extensive quadruple-$Q$ manifold found here provides a concrete example of how quartic mode coupling can generate a hierarchy of distinct noncoplanar spin crystals with tunable scalar spin chirality landscapes under an external magnetic field.
Our work thus offers a systematic platform for exploring and classifying higher-order multiple-$Q$ phases in square-lattice metallic systems and related engineered structures where competing RKKY interactions and higher-order couplings are expected to be operative.

\section*{Acknowledgement}
This research was supported by JSPS KAKENHI Grants Numbers JP22H00101, JP22H01183, JP23H04869, JP23K03288, JP23K20827, and by JST CREST (JPMJCR23O4) and JST FOREST (JPMJFR2366).


\bibliographystyle{cmpj}
\bibliography{ref}

%
%

\ukrainianpart

\title{Налаштовані полем чотири-$Q$ фази, зумовлені імпульсно-просторовою фрустрацією}
\author{С. Хаямі}
\address{
	Вища школа наук, Університет Хоккайдо, Саппоро 060-0810, Японія
}

\makeukrtitle

\begin{abstract}
	\tolerance=3000%
Кратний $Q$-магнетизм у системах вільних електронів дозволяє утворювати складні спінові кристали та некомпланарні текстури навіть у центрально-симетричних полях. Ми вивчаємо примітивну спінову модель в імпульсному просторі на квадратній ґратці з чотирма хвильовими векторами впорядкування, пов'язаними з симетрією, включаючи білінійну та біквадратну взаємодії під дією позаплощинного магнітного поля. Використовуючи імітацію відпалу, ми отримуємо залежну від поля фазову діаграму та ідентифікуємо послідовні переходи між одинарним $Q$, подвійним $Q$ та кількома нееквівалентними кратними $Q$-станами. Кратний $Q$ многовид демонструє багату внутрішню структуру: стани, що мають однакові хвильові вектори, відрізняються фазовою синхронізацією, розподілом амплітуди та некомпланарністю, що призводить до різних просторових текстур та появи скалярних спінових хіральних структур. Наші результати показують, що фрустрація в імпульсному просторі та біквадратна взаємодія забезпечують ефективний шлях до стабілізації різноманітних кратних $Q$-спінових кристалів, даючи загальну основу для появи спінових текстур вищого порядку в центросиметричних магнітах з вільними електронами.
	\keywords кратні $Q$-стани, квадратна ґратка, біквадратна взаємодія, фрустрація в імпульсному просторі, взаємодія РККІ, центрально-симетричні магніти
	
\end{abstract}

  \lastpage
  \end{document}